\documentclass[preprint]{aastex}
 
\usepackage{amssymb}
\usepackage{epsfig}
\usepackage{amsmath}
\usepackage{subfigure}
 
\begin{document}

\title{Stellar Encounters with the $\beta$ Pictoris
Planetesimal System}

 
\author{Paul Kalas\altaffilmark{1,2}, Jean-Marc Deltorn\altaffilmark{1,3}}
\affil{Space Telescope Science Institute,
Baltimore, MD 21218
}
\and 
\author{John Larwood\altaffilmark{4}}
\affil{Queen Mary \& Westfield College, London, England}
\email{kalas@astron.berkeley.edu}
 
 
\altaffiltext{1}{Space Telescope Science Institute, 
3700 San Martin Drive, Baltimore, MD 21218
}
\altaffiltext{2}{Current Address:  
Department of Astronomy, 601 Campbell Hall, University of California,
    Berkeley, CA 94720}
\altaffiltext{3}{Laboratoire d'Astrophysique de Marseille, 
Traverse du Siphon, Les trois-Lucs, 13376, Marseille Cedex 12, France
}
\altaffiltext{4}{{Astronomy Unit, Queen Mary \& Westfield College, London, E1 4NS, United Kingdom}
}

\begin{abstract}
We use data from the {\it Hipparcos} catalog and the \citet{bbf00} catalog
of stellar radial velocities to
test the hypothesis that the $\beta$ Pic planetesimal disk
was disrupted by a close stellar encounter.
We trace the space motions of 21,497 stars and 
discover 18 that have passed within 5 pc of $\beta$ Pic in the
past 1 Myr.
$\beta$ Pic's closest encounter is with the
K2III star HIP 27628 ($\sim$0.6 pc), but dynamically the most
important encounter is with the F7V star
HIP 23693 ($\sim$0.9 pc).  
We calculate the velocity and eccentricity changes induced
by the 18 perturbations and conclude that they are dynamically 
significant if planetesimals exist in a $\beta$ Pic Oort
cloud.
We provide a first-order estimate for the evolutionary
state of a $\beta$ Pic Oort cloud and conclude that the primary role
of these stellar perturbations would be to help build a comet
cloud rather than destroy a pre-existing structure.
The stellar sample 
is $\sim$20\% complete and motivates future
work to identify less common close interactions that 
would significantly modify the observed circumstellar disk.
For future radial velocity study we identify 6 stars 
in the {\it Hipparcos} catalog that may have approached $\beta$ Pic
to within 0.1 pc and therefore remain as candidate disk perturbers.

\end{abstract}
 
 

\section{Introduction}

The dynamical mechanisms that dominate the formation and subsequent 
evolution of planetary systems may be broadly 
described as either endogenic or exogenic in origin.  
The formation of giant planets, and their apparent 
orbital migration \citep{gol80,mal00, mar00} 
are probably the most significant endogenic source for 
modifying a planetary system over time.  In our solar system,
the giant planets have displaced a significant fraction of
small bodies from their formation sites to either the 
Oort cloud or interstellar space.
For circumstellar 
dust disks observed around nearby main sequence stars, unseen giant 
planets are believed to produce the ubiquitous central 
depletions in the dust distributions \citep{roq94, 
pan97,gre98,liu99,wya00}.

However, young systems, such as pre-main sequence stars
in clusters and the proplyd objects 
in Orion, experience significant exogenic forces \citep{lar97}.  
Close stellar encounters with other cluster members \citep{lau98} and radiation 
pressure from nearby massive stars \citep{hen99}
remove disk material 
and can disrupt the vertical settling of circumstellar 
dust and gas.  During the 
evolution of our Solar System, the galactic tide and encounters with molecular 
clouds and passing stars
decoupled comets from the planetary
region, preserving them in
the Oort cloud 
\citep{oor50,fer97}.

The dust disk around the A5V star $\beta$ Pic displays evidence for both types
of dynamical mechanism.  A planet may be responsible for clearing dust within 
$\sim$20 AU radius of the star \citep{smi84,lag94}, 
perturbing families of comets towards the stellar photosphere \citep{beu00},
vertically 
disrupting the disk at 50 AU radius \citep{bur95,mou97,hea00},
and creating a few-hour duration, 0.06 magnitude, 
achromatic drop in $\beta$ Pic's lightcurve \citep{lec97}.
A recent and close stellar flyby perturbation 
may be responsible for producing the radially 
and vertically asymmetric disk structure at large radii 
\citep{kal95}, as well as 
substructure in one disk midplane between
500 and 800 AU projected radius \citep{kal00,lar00}. 
Since the age of $\beta$ Pic is
between 8 Myr \citep{cri97}
and 20 Myr \citep{bar99}, understanding the dynamics of objects
surrounding $\beta$ Pic could elucidate 
conditions and events that determined the early
evolution of our Solar System.

Here we will examine possible exogenic perturbations on $\beta$ Pic.
A key piece of observational evidence needed to confirm
the stellar flyby hypothesis is to identify a perturbing star.
Given improved trigonometric parallaxes 
and proper motions from {\it Hipparcos} \citep{esa97}, 
we are 
motivated to test the stellar flyby hypothesis empirically. 
Section 2 presents our method for 
using $Hipparcos$ and radial velocity data to 
trace the galactocentric motions of
stars relative to $\beta$ Pic.  Section 3 
identifies stars that
have passed within 5 pc of $\beta$
during the past 1 Myr.
In Section 4 we 
assess the significance
of the stellar encounters for the evolution
of a possible Oort cloud around $\beta$ Pic.
Section 5 demonstrates the plausibility of
very close stellar encounters and presents a list of candidate perturbers
for future radial velocity observations.

\section{Search for $\beta$ Pic stellar perturbers}

We search for candidate $\beta$ Pic perturbers using data from the entire 
{\it Hipparcos}
catalog and the \citet{bbf00} catalog of stellar radial velocities.
The {\it Hipparcos} catalog gives the positions, proper motions
and parallaxes of 118,218 stars from which we compute the galactocentric
coordinates.
\citet{bbf00} provide the mean radial velocities for 36,145 stars, of which
21,497 stars are also contained in the {\it Hipparcos} catalog.  We combine the
radial
velocities with the $Hipparcos$ proper motion data \citep{joh87} to
trace the space trajectories of the 21,497 stars during the past 1 Myr.

The observed disk substructure should be short-lived
due to orbital phase mixing.
The dynamical models of \citet{kal00} and \citet{lar00} indicate that
the perturbation occurred as recently as 10$^5$ yr ago.  However, we choose
to extend the search 1 Myr into the past to allow for the uncertainties introduced by the
assumptions inherent in the models.
 
Key factors limiting the completeness of the search are the sensitivity
constraints imposed by the two catalogs. 
In general, we expect that a significant fraction of late type stars near 
$\beta$ Pic ($d$ = 19.3 pc) is undetected.
For spectral types later than G5, Volume 1 of the $Hipparcos$ catalog states
a completeness limit:  $m_v\le 7.3 + 1.1 |$sin$(b)|$, where $b$ is the
galactic latitude.  For $\beta$ Pic, $b\sim-30$ degrees, giving a completeness limit
$m_v\le 7.85$ mag.  At $d$ = 20 pc
this translates
to $M_v \le$ 6.3 mag, which corresponds to the absolute magnitude of a K2V star 
\citep{cox99}.
At $d$ = 30 pc, $M_v \le 5.5$ mag,
which is the absolute magnitude of a K0V star \citep{cox99}.  
Thus, the $Hipparcos$ catalog 
includes a significant fraction of the A - G spectral types in a 10 pc
radius volume around $\beta$ Pic, but will 
miss K and M dwarfs  [in \citet{kal00} the 
$\beta$ Pic disk perturber has a mass consistent with an M dwarf].

Estimating the completeness of the \citet{bbf00} catalog is considerably more
difficult because it is not an all-sky survey, and the radial velocity information
derives from different sources.  
Even though the majority of stars are $m_v\le 10$ mag, comparable to
the sensitivity limit of the $Hipparcos$ catalog, the $Hipparcos$ 
catalog contains more than
three times as many stars. 
The incompleteness of our sample
therefore exceeds 50\%, and in Section 3 we estimate
an incompleteness of $\sim$80\%.  
Future missions such
as $FAME$, $SIM$ and $GAIA$ will have the sensitivity to
significantly improve the perturber search if
radial velocity catalogs are also
expanded.

For $\beta$ Pic's radial velocity we choose the recent
measurement given by \citet{gre99} because the published
error, $\sigma_{RV}$, is smaller than that determined by \citet{bbf00}.  
Experimental trials using other published
radial velocities for $\beta$ Pic (e.g. Lagrange et al. 1995) show
negligible differences from the results presented below.

The relatively short timescale considered here permits 
a first order estimate of the stellar trajectories using a straight line motion
approximation.  
A correction for the two body
interaction  
is only significant 
for very close 
($\lesssim 10^{-2}$ pc) and low relative velocity ($\lesssim 10$ km s$^{-1}$) encounters. 
For each of the 21,497 stars we calculate a trajectory backward in time 
and determine the closest approach distance, $D_{ca}$ (pc), to $\beta$ Pic and the 
time of closest approach, $t_{ca}$ (kyr), where $t_{ca}$ = 0 is the present.
To test the validity of the linear approximation for deriving trajectories we 
use a fourth order Runge-Kutta scheme to solve the equation of motion 
in the Galactic potential for six of the 18 stars
that approach within 5 pc of $\beta$ Pic.  
We adopt the
axisymmetric Milky Way mass distribution model provided by 
\citet{als91}. This model\footnote{Consisting of a spherical central bulge, a 
\citet{miy75} disk, and a massive spherical halo.} has 
proven to be well adapted to the 
calculations of orbits \citep{deb97} and has been already applied to the 
motions of globular clusters and nearby stars \citep{als93,sch97}. 
In the galactocentric cylindrical coordinates ($R,\theta,z$)
the small value of $z/R$ in
the range of time considered here permits the 
decoupling of motions in the galactic plane and on the $z$ axis.
We check our calculations by reproducing the published results on 
the nearest approaches of stars with the Sun 
\citep{gar99,mul96,mat94}. 
The initial coordinates and velocities have been changed from the heliocentric 
to the galactocentric 
reference frame using the current IAU values for the LSR 
($\dot{\theta}_{LSR}$=220 km s$^{-1}$, R$_{LSR}$=8.5 kpc).
As expected, the agreement between the ($D_{ca},t_{ca}$) values derived from 
the straight line approximation and the ones calculated from the integrated 
Galactic orbits allows us to restrict our computations to the linear case for 
the selection of candidates.   

The main advantage of adopting the 
the linear approximation is that we can determine 
the influence of the errors for the input positions and velocities on
the estimate of ($D_{ca},t_{ca}$) through a
Monte-Carlo draw.  We use the standard deviations
provided in the {\it Hipparcos} catalog and the \citet{bbf00} catalog.
We assume a Gaussian distributions for the errors,
centered on the average value and
with dispersion equal to $\sigma_{rms}$. We produce 10$^4$ random draws
that for each star lead to a probability distribution in the
closest approach plane ($D_{ca},t_{ca}$). 

Table 1 lists the stars that approached within 5 pc of $\beta$
Pic during the past 1 Myr.  For this subset of stars we then use $SIMBAD$ to search the 
literature for recent radial velocity
measurements that may not have been included in the
\citet{bbf00} catalog.
\citet{gre99} have measured the radial velocity
for HIP 19893 and give a smaller measurement error than that listed
by \citet{bbf00}.  We therefore re-calculate the space trajectory
and errors using the \citet{gre99} radial velocity, which is
given in Table 1. 
We make the same correction for HIP 17378, which
has a more accurate radial velocity measurement from \citet{dem99}.
As with $\beta$ Pic, trials using 
different published radial velocity measurements
for each star
produce negligible changes in the final values of $D_{ca}$ and $t_{ca}$.

\section{Candidate perturbers with ${D}_{ca} <$ 5 pc}

From the 21,497 stars selected from the {\it Hipparcos}
catalog and \citet{bbf00} catalog, 
18 are found to have  ${D}_{ca} < 5$ pc with respect
to $\beta$ Pic in
-10$^6$ yr $<t_{ca}<$ 0.   
Table 1 lists the values of their respective trajectory parameters.
Columns 2 and 3 in Table 2 give the maxima and 1$\sigma$ uncertainties for the
probability density distributions derived from the Monte-Carlo draws for
(${D}_{ca},{t}_{ca}$).
Figure 1 shows the $D_{ca},t_{ca}$ values resulting from
the Monte Carlo draw for a subset of stars with ${D}_{ca} <$ 3 pc.  
Figure 2 presents the results
in the form of isocontours representing
the 68.3\%, 95.4\%, and 99\% confidence levels for finding a star within a given
region in the close encounter plane ($D_{ca},t_{ca}$).
An encounter probability
is determined by dividing
the contour area below a specified closest approach limit,
by the total contour area in the $D_{ca},t_{ca}$ plane. 
The probabilities 
that each star approached $<$1 pc and $<$ 0.5 pc with confidence
levels of  95.4\% and 68.3\% 
are listed in columns 5 and 6 in Table 2.

Three stars - HIP 23693, HIP 27628, and HIP 29958 -
have ${D}_{ca}\leq$ 1 pc.
However, the greater uncertainties in the observables 
for HIP 29958 (Table 1) mean that there is only a 10\% probability at
the 95.4\% confidence level that its closest approach was $<$1 pc.
For HIP 23693 and HIP 27628, the probabilities that 
$D_{ca}<$ 1 pc are $>$50\%.
HIP 93506 and HIP 116250 are notable because 2-3$\times 10^5$ yrs before their
closest approach with $\beta$ Pic (Table 2), 
they passed $<$3 pc from the Sun \citep{gar99}.

The incompleteness of our sample is evident by comparing the 
number of candidate perturbers found by this experiment to the number expected.
If we assume that $\beta$ Pic's heliocentric distance, $d$ = 19.3 pc, places 
it within the Solar neighborhood, then the stellar encounter frequency
should be roughly equal to that of the Sun:
$N = 12.4$ Myr$^{-1}$ 
for $D_{ca}\leq$1 pc \citep{gar99}.
Thus, our search produces a factor of $\sim$5
too few perturbers.  Using the $Hipparcos$ catalog, \citet{gar99} give
their empirical finding that $N\geq$ 3.5 D$^{2.12}_{ca}$ Myr$^{-1}$, and
they conclude that their sample is incomplete by at least 50\%.
For our $D_{ca} < 5$ pc cut-off, the empirical relation given by
\citet{gar99} yields $N \geq$ 106 Myr$^{-1}$, but for $\beta$ Pic we find
$N = 18 $ Myr$^{-1}$ (Table 1). Again, the discrepancy indicates that our sample
is $\sim$20\% complete.  Thus we expect that a volume-limited search realized
with future stellar catalogs will reveal at least $\sim$10$^2$ stars approaching
$\beta$ Pic within 5 pc.

To determine which candidate perturber had the greatest dynamical
impact on $\beta$ Pic, we 
factor in the relative stellar masses and velocities. 
We calculate
the velocity impulse $\Delta v$ due to each stellar 
passage on both a $\beta$ Pic 
disk particle ($r$=10$^3$ AU) 
and on a hypothetical Oort cloud object ($r$=10$^5$ AU). 
Under the impulse approximation, the change of velocity $\Delta v$ 
of a comet relative to $\beta$ Pic due to the influence of a passing 
star can be approximated as,
\begin{equation}
\Delta v^2=\biggl(\frac{2{\mathcal{G}}M_*}{{\Delta}V_{ca}D_{ca}}\biggr)^2\frac{r^2}{r^2+D_{ca}^2-2rD_{ca}cos\beta}
\end{equation}
where  $M_{\star}$ is the mass of the 
passing star, ${\Delta}V_{ca}$ is 
the relative velocity, and $D_{ca}$ and $r$ are 
the distances from $\beta$ Pic 
to the passing star and to a comet, respectively. 
$\beta$ denotes the angle between 
$\mathbf{r}$ and $\mathbf{D}_{ca}$ at the time of closest 
approach \citep{oor50,fip91}.

From ${\Delta}v$ we can also estimate the change
in eccentricity, ${\Delta}e$,  using the following result
from numerical simulations \citep{bru96}:  
${\Delta}e\sim2{\Delta}v/v$, where the comet's orbital velocity
$v$ is $v^2= \mathcal{G}M_{\beta Pic}/r$.
Integrating Eq. 1 over $\beta$ we compute the average perturbation
of a given close encounter on a shell of comets at distance $r$.
The resulting ${\Delta}v$ and ${\Delta}e$ for both the maximum perturbation 
($\beta=0$) and the average one at $r=10^5$ AU are compiled 
in Table 3, together with the 1 $\sigma$ uncertainties deduced 
from our Monte-Carlo draws, and plotted in Figures 3a, 3b, 4a, and 4b.  
For $r=10^3$ AU, the values for ${\Delta}v$ and ${\Delta}e$ are smaller
by a factor determined from Eqn. 1.  For example, with the HIP 23693 encounter,
${\Delta}v_{max}$ is a factor of $4.4 \times 10^4$ smaller, and ${\Delta}e_{max}$ is
$4.4 \times 10^5$ smaller.

Among the initial best candidates,
HIP 27628 (\#8) produces the greatest ${\Delta}v_{max}$ on particles located between
itself and $\beta$ Pic (${\Delta}v_{max}\simeq$ 0.7 m s$^{-1}$) at the 
time of closest approach (Table 3). However, its
effect on the whole cloud is diminished due to its small mass and large relative 
velocity with respect to $\beta$ Pic.
Similarly, HIP 29958 (\#10), with $M_*\sim0.5 M_{\odot}$ and 
${\Delta}V_{ca}=$102.1 km s$^{-1}$,
does not induce on average any notably large 
perturbation over the orbiting comets, even at $r=10^5$ AU.
HIP 23693 (\#6)  remains then as the most significant perturber 
as it combines the smallest ${D}_{ca}$ in our sample, a rather 
small relative velocity to  $\beta$ Pic (22$\pm 3$ km s$^{-1}$) and a 
large mass ($\sim 1.4 M_{\odot}$). 

\section{Dynamical influence on the evolution of a $\beta$ Pic Oort cloud}

The candidate stellar perturbers detected by our search cannot
account for the
$\sim$0.005 pc close encounter simulated by \citet{kal00} to explain
the observed dust disk morphology.  
However, the values for ${\Delta}v$ and ${\Delta}e$ at $r=10^5$ AU
may be large enough to impact the dynamical evolution
of objects located beyond the detected dust disk, in a $\beta$ Pic analog
to the Solar Oort cloud.  For a circular orbit at $r=10^5$ AU, the escape
velocity for a $\beta$ Pic comet, 176 m s$^{-1}$, is significantly greater
than the orbital velocity, 124 m s$^{-1}$.  Thus, only the
cumulative effect of perturbers will dynamically modify a $\beta$ Pic
Oort cloud.  On the other hand, the values for ${\Delta}e$ are large
enough that objects already on eccentric orbits, $e > 0.99$, could be
stripped from the system after a single stellar perturbation.  
Moreover, objects with $r>10^5$ AU will
experience a stronger perturbation that could lead to ejection.
Below we determine how the $D_{ca}$ values for the candidate perturbers compare 
to the radius of a $\beta$ Pic Oort cloud, and then discuss
the possible evolutionary effects of the perturbations.

The maximum size of a
planetesimal cloud  gravitationally bound to $\beta$ Pic
is defined by the Roche surface of the star, $a_{t}$,  
set by the galactic tidal field
\citep{tre93}:
\begin{equation}
a_t = 1.7\times 10^5 AU \biggl(\frac{M_{\star}}{M_{\odot}}\biggr)^{1/3}
\biggl(\frac{\rho}{0.15 M_{\odot} pc^{-3}}\biggr)^{-1/3}
\end{equation}
\citet{hol00} give the variance-weighted average from seven different
studies for the galactic mass density, $\rho$ = 0.11 $M_{\odot} pc^{-3}$.
If $M_{\beta Pic}$ = 1.75 $M_{\odot}$ \citep{cri97}, then
$a_t = 2.2 \times 10^{5}$ AU, or $\sim$1.1 pc.  
Thus, at least three of the candidate perturbers (Table 2) penetrate 
$\beta$ Pic's Roche radius.

Other perturbers, though outside the Roche radius, 
pass close enough to temporarily exceed the 
gravitational influence of the galactic tidal field.
The distance from $\beta$ Pic where the gravitational
forces of $\beta$ Pic and the perturber on an Oort cloud comet 
are equal is given by \citep{mul96}:
\begin{equation}
R_{eq} = 
D_{ca}[1-(1+\sqrt{M_{\beta Pic}/M_{*}})^{-1}]
\end{equation}
The last column of Table 3 gives $R_{eq}$ for each of the candidate
perturbers.  In addition to the three perturbers that physically enter the Roche radius,
HIP 25544 has $R_{eq} < a_t$.
Thus a total of four stars 
out of the 18 in Tables 1 and 2 
penetrate a possible Oort cloud around $\beta$ Pic. 

In general, stars passing near an Oort cloud will
either destroy a fraction  of the cloud by sending comets 
into interstellar space or
closer to the central star, or help build the cloud by increasing
the periastron distances and hence the dynamical lifetimes of comets 
in the presence of planets. 
Stellar passages within an Oort cloud may induce 
comet showers which may briefly increase dust replenishment near the star \citep{wei96}.
To determine which outcome results from the
stellar perturbations identified here,
we must first consider if the existence of a $\beta$ Pic Oort cloud is
plausible, and if so estimate its evolutionary state. 

The four basic conditions for creating an Oort cloud around a star are 
(e.g. Fernandez 1997): 1) the
formation of planetesimals in a region influenced by planets,
2) the existence of a massive planet
to dynamically pump the semi-major axes of planetesimals from their
formation site to large distances, 3) exogenic
perturbations that decouple the planetesimals from the planets by
increasing the planetesimals' periastra, and 4) sufficient time
(stellar age $>$ relevant timescales).

For the first condition,
the existence of meter to kilometer sized planetesimals around $\beta$ Pic is inferred from:
a) the short lifetime of dust particles relative to $\beta$ Pic's 
stellar age that implies a source of replenishment, probably from the collisional
erosion of larger, unseen parent bodies \citep{bap93}, and b) variable,
transient absorption features modeled as the rapid sublimation of
comet-like bodies near the photosphere \citep{beu00}.  
For the second condition,
indirect
evidence for 
a massive planet at $\sim$1-10 AU radius is summarized in Section 1.
The third condition, exogenic perturbations, are expected from passing
stars, molecular clouds, and the galactic tidal field.

For the fourth condition, we compare $\beta$ Pic's age, $t_{\beta Pic}$, 
to the timescales required to produce
an Oort cloud, as quantified by
\citet{tre93}.
Two of the most important timescales are the diffusion
timescale, $t_{d}$, and the freezing timescale, $t_{f}$.
The diffusion timescale is the time for
comet apastra to diffuse
out to the Roche radius of a star at constant periastra
(i.e. the time for a comet's energy to reach the
escape energy from repeated interactions with a planet during periastron).
The freezing timescale is the time for the galactic tide
to increase comet periastra beyond the planet region, thereby
freezing any further increase in the apastra.  
For a mature
Oort cloud (i.e. well-populated and dynamically long-lived)
around $\beta$ Pic we require  $t_{f} \leq t_{d} < t_{\beta Pic}$.

If $t_{f} > t_{d}$,
then comet
apastra will grow too quickly and achieve escape energy with one
final encounter with a planet.
The timescales are therefore sensitive to planet mass, $M_p$, 
and semi-major axis, $a_p$.
If $t_{\beta Pic} < t_{d}$, then the
system may be in the process of forming an Oort cloud, but the comets
are probably still coupled to the planetary region.  

To satisfy  $t_{d} < t_{\star}$, \citet{tre93} gives the 
following expression:
\begin{equation}
\frac{M_p}{M_{\oplus}}\geq\biggl(\frac{M_{\star}}{M_{\odot}}\biggr)^{3/4}
\biggl(\frac{t_{\star}}{10^{9}yr}\biggr)^{-1/2}
\biggl(\frac{a_p}{1 AU}\biggr)^{3/4}
\end{equation}
For $t_{f} \leq t_{d}$, the following relation must be satisfied:
\begin{equation}
\frac{M_p}{M_{\oplus}}\leq 1.7\biggl(\frac{M_{\star}}{M_{\odot}}\biggr)^{5/7}
\biggl(\frac{\rho}{0.15 M_{\odot} pc^{-3}}\biggr)^{2/7}
\biggl(\frac{a_p}{1 AU}\biggr)^{6/7}
\end{equation}
The age-independent constraint given by Eqn. 5 is significant because
it generally implies that no dynamically stable Oort cloud will form 
if $M_p \geq M_{Saturn}$,
even for a wide range of stellar masses.  For example the Saturn-mass
extrasolar planet candidates around HD 16141 and HD 46375 \citep{mbv00} will
not produce an extrasolar Oort cloud. 

In Figure 5 we plot the regions that satisfy  $t_{f} \leq t_{d} < t_{\beta Pic}$ for
a range of
$M_p$ and $a_p$, given $M_{\beta Pic}/{M_{\odot}}=1.75$, 
$\rho$ = 0.11 $M_{\odot} pc^{-3}$ and $t_{\beta Pic}$ = $10^7$, $10^8$ and $10^9$ yr.
For $\beta$ Pic's estimated age $\sim$10$^7$ yr,
there is no combination of $M_p$ and $a_p$ that is capable of producing 
an Oort cloud by the present epoch
that is stable from ejection.  In fact, a stable Oort cloud
is possible only towards the end
of $\beta$ Pic's lifetime, $\sim$10$^9$ yr.
If we assume a combination of $M_p$ and $a_p$
that will eventually result in a $\beta$ Pic Oort
cloud (e.g. $M_p$ = 10 M$_{\oplus}$, $a_p=5$ AU), then at the present epoch  
comet apastra are only a few
percent of their final values at $\sim$10$^5$ AU.  
The dust disk is detected in scattered light 
as far as $\sim2\times10^3$ AU \citep{lar00}, or
$\sim$1$\%$ of the Roche radius.  Depending on
how future observations refine our
knowledge of a planetary system around $\beta$ Pic,
the observed dust disk may trace the early evolution of
an extrasolar Oort cloud.

In summary, objects orbiting $\beta$ Pic near $10^5$ AU radius
might still be coupled to $\beta$ Pic's planetary region.  
The galactic tidal field is insufficient to decouple these objects from
the planetary region before they attain escape energy.  
The stellar perturbations identified in our search
neither eject a significant number of comets into interstellar space,
nor do they cause comet showers near the star.
The evolutionary role of these stellar perturbations 
is to build $\beta$ Pic's
Oort cloud
by decoupling comet periastra from the planetary region.

\section{Candidate perturbers for future radial velocity study}

The most significant constraint on the completeness of this
search is the lack of radial velocity information for
nearly 100,000 stars in the $Hipparcos$ catalog.  
The completion of an all-sky radial velocity survey,
such as one proposed for the $GAIA$ mission, has the potential to
dramatically increase our knowledge of galactic
space motions.  

However, given only the position, distance and proper motion data
from the $Hipparcos$ catalog, it is possible to exclude the
stars that would not approach $\beta$ Pic for any physically meaningful value
of radial velocity, and identify a sample of stars
that could plausibly encounter $\beta$ Pic.
We adopt an iterative approach, 
first
assuming -120 km s$^{-1}<R_V<$ 120 km s$^{-1}$, $\Delta R_V$=10 km s$^{-1}$,
to calculate $D_{ca}$ and $t_{ca}$ during the past 2 Myr.  
We select the stars with $D_{ca}<$ 5 pc,
and repeat the procedure using $\Delta R_V$= 1 km s$^{-1}$, and
finally $\Delta R_V$= 0.1 km s$^{-1}$ for the stars with
$D_{ca}<$ 1 pc. We fix the initial positions, 
distances, and proper motions
to the mean values given by the $Hipparcos$ catalog,
rather than utilize the Monte Carlo
method discussed in Section 2. 
 
The results are displayed in Tables 4 and 5.  Tables 4
gives the maximum number of perturbers
that can be ``produced'' by adjusting the radial velocity in order to
get the closest possible approach to $\beta$ Pic. 
A maximum of 93 stars 
approach $\beta$ Pic with
$D_{ca}<$ 1 pc and $t_{ca}>$ -1 Myr.
As expected, very close and very recent crossing
events are rare, and the more the
time and distance constraints are relaxed, the more events 
are produced.
The values 
given in Tables 4 are upper limits to the number of 
encounters experienced by $\beta$ Pic, exceeding the
expected rate of $<$1 pc encounters per Myr by
a factor of 7.5 (Section 3). 

Table 5 gives the 22 perturber candidates with $D_{ca}<0.5$ pc
during the past 1 Myr. The columns designate the:
(1) name of the star, 
(2) distance of the closest possible approach, 
(3) corresponding time,
(4) radial velocity that produces $D_{ca}$,
(5) interval of radial velocities such that the distance of closest 
approach remains below 0.5 pc,
and
(6) relative velocity at the closest possible approach.
Six candidate perturbers (marked with an asterisk) have $D_{ca}<$ 0.1 pc and 
are capable of perturbing the 
observed dust disk.    
Table 5 both gives a specific list of objects that
require follow-up observations to determine their
true radial velocities, and demonstrates 
the plausibility of very
close stellar encounters with $\beta$ Pic.

\section{Summary}
We have tested the hypothesis that $\beta$ Pic
experienced a close stellar flyby by tracing the
space motions of 21,497 stars 1 Myr into the past.
We used the $Hipparcos$ catalog for position, proper motion
and parallax data, and the \citet{bbf00} catalog
for radial velocity data. The completeness of our sample
is $\sim$20\% and is most
sensitive to B-G spectral types.  Our findings are:

1) In the past 1 Myr, 18 stars in our sample have had a closest approach
to $\beta$ Pic ${D}_{ca}<$ 5 pc.  

2) Two stars, HIP 23693 and HIP 27628, have $>$50\% probability 
that ${D}_{ca} <$ 1 pc.  The probability 
that
${D}_{ca} <$  0.01 pc is negligible,
meaning that these encounters
cannot account for the large-scale $\beta$ Pic disk asymmetries
modeled by \citet{kal00} and \citet{lar00}.

3)  The most significant dynamical perturbation on the $\beta$ Pic system is due to
the HIP 23693 encounter $\sim$356 kyr in the past.  Averaged over a hypothetical
$\beta$ Pic Oort cloud 0.5 pc in radius, the encounter induced a mean
velocity change $\simeq$ 0.3 m s$^{-1}$, 
and eccentricity change $\simeq$0.005.  

4)  The Roche radius of $\beta$ Pic set by the galactic tidal field is
$\sim$1.1 pc.  Four stellar perturbers
penetrate the Roche radius in the past 1 Myr and could
dynamically influence planetesimals bound to $\beta$ Pic.

5)  We summarize evidence favoring the formation of a $\beta$ Pic Oort cloud.
However, we find that $\beta$ Pic is probably too young to have an Oort cloud 
that is decoupled from the planetary region.  The stellar perturbations
are significant in helping $\beta$ Pic build its
Oort cloud by pumping comet periastra away from dynamically unstable regions
near planets.  

6)  We identify a sample of 22 stars detected by $Hipparcos$ that are candidate
$\beta$ Pic perturbers, but require future 
radial velocity observations to determine their closest approach distances.
Six of these stars may have $D_{ca}<0.1$ and are potential disk perturbers.

\acknowledgments
 
{\bf Acknowledgements:} The authors are grateful to AURA and the
Space Telescope Science Institute for support.
This research has made use of the SIMBAD database and the VizieR service,
operated at CDS, Strasbourg, France.


\pagestyle{headings}
\begin{table}
\begin{center}
\epsfbox{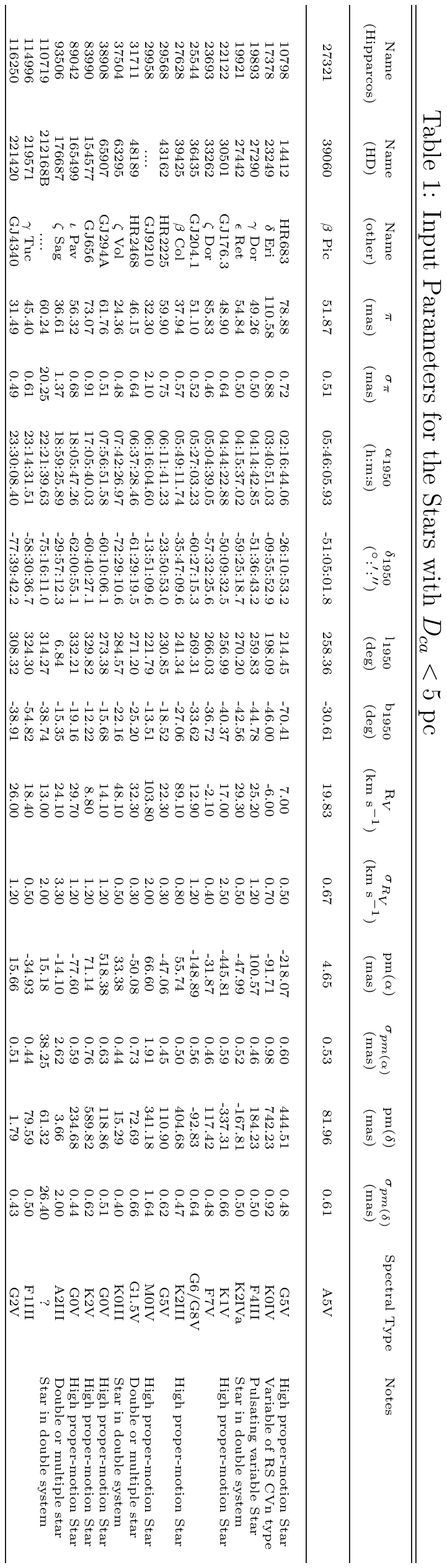}
\caption{}
\end{center}
\end{table}

\begin{table}
\caption{Characteristics of the Closest Approach for the Stars with ${D}_{ca}<5$pc}
\begin{tabular}[h]{cccccc}
\hline \hline\\
$\#$&Name            & t$_{ca}$  & ${D}_{ca}$      & P$_{95.4\%}(<1pc)$        & P$_{95.4\%}(<0.5pc)$      \\
\vspace{0.2cm}
&(Hipparcos)     & (kyr)         & (pc)          & (P$_{68.3\%}(<1pc)$) ($\%$) & (P$_{68.3\%}(<0.5pc)$) ($\%$) \\\hline
\\
\vspace{0.2cm}
1&HIP 10798  &-318.2$^{+11.7}_{-14.6}$& 1.88$^{+0.51}_{-0.40}$&  2.49 (0.00)              & 0.00 (0.00) \\
\vspace{0.2cm}
2&HIP 17378  &-294.7$^{+16.6}_{-17.9}$& 3.96$^{+0.42}_{-0.41}$&  0.00 (0.00)              & 0.00 (0.00) \\
\vspace{0.2cm}
3&HIP 19893  &-31.1$^{+35.2}_{-36.3}$ & 4.94$^{+0.19}_{-0.14}$&  0.00 (0.00)              & 0.00 (0.00) \\ 
\vspace{0.2cm}
4&HIP 19921  &-118.9$^{+11.7}_{-8.9}$ & 3.74$^{+0.31}_{-0.33}$&  0.00 (0.00)              & 0.00 (0.00) \\ 
\vspace{0.2cm}
5&HIP 22122 & -34.0$^{+5.9}_{-5.8}$   & 2.76$^{+0.56}_{-0.18}$&  0.00 (0.00)              & 0.00 (0.00) \\ 
\vspace{0.2cm}
6&HIP 23693  &-356.3$^{+29.3}_{-32.2}$& 0.92$^{+0.13}_{-0.12}$&  54.16 (53.42)    & 0.00 (0.00) \\ 
\vspace{0.2cm}
7&HIP 25544  &-116.0$^{+8.8}_{-9.1}$  & 1.49$^{+0.39}_{-0.30}$&  0.00 (0.00)              & 0.00 (0.00) \\ 
\vspace{0.2cm}
8&HIP 27628  &-107.2$^{+14.6}_{-11.7}$& 0.58$^{+0.51}_{-0.11}$&  70.28 (71.85)    & 3.10 (2.75) \\
\vspace{0.2cm}
9&HIP 29568  &-693.2$^{+43.9}_{-46.9}$& 2.95$^{+0.90}_{-0.32}$&  0.00 (0.00)              & 0.00 (0.00) \\
\vspace{0.2cm}
10&HIP 29958  &-198.0$^{+32.2}_{-29.3}$& 1.00$^{+2.36}_{-0.40}$&  10.21 (12.76)    & 0.70 (0.00) \\ 
\vspace{0.2cm}
11&HIP 31711  &-189.3$^{+49.8}_{-76.2}$& 3.97$^{+0.37}_{-0.31}$&  0.00 (0.00)              & 0.00 (0.00) \\ 
\vspace{0.2cm}
12&HIP 37504  &-643.4$^{+52.7}_{-38.1}$& 4.59$^{+1.52}_{-0.91}$&  0.00 (0.00)              & 0.00 (0.00) \\ 
\vspace{0.2cm}
13&HIP 38908  &-139.7$^{+5.9}_{-8.7}$      & 1.97$^{+0.69}_{-0.42}$&  0.00 (0.00)          & 0.00 (0.00) \\ 
\vspace{0.2cm}
14&HIP 83990  &-303.5$^{+17.6}_{-14.6}$& 3.92$^{+1.18}_{-1.16}$&  0.00 (0.00)              & 0.00 (0.00) \\
\vspace{0.2cm}
15&HIP 89042  &-397.3$^{+20.5}_{-23.4}$& 2.29$^{+1.21}_{-0.21}$&  0.00 (0.00)              & 0.00 (0.00) \\
\vspace{0.2cm}
16&HIP 93506  &-910.0$^{+137.7}_{-128.1}$& 3.50$^{+2.31}_{-1.21}$&  0.00 (0.00)            & 0.00 (0.00) \\
\vspace{0.2cm}
17&HIP 114996 &-596.5$^{+32.6}_{-42.7}$& 3.94$^{+1.35}_{-1.77}$&  0.00 (0.00)              & 0.00 (0.00) \\ 
\vspace{0.2cm}
18&HIP 116250 &-883.6$^{+61.5}_{-67.4}$& 2.79$^{+2.39}_{-1.92}$&  3.66 (2.74)              & 0.08 (0.00) \\
\hline
\end{tabular}
\end{table}
 
\begin{table}
\scriptsize
\caption{Dynamical Influence of the Close Encounters on Objects Orbiting $\beta$ Pic at $10^5$ AU}
\begin{tabular}[h]{lcccccccccc}
\hline \hline\\
$\#$ & Name     & Spectral& Mass        &       D$_{ca}$        &  ${\Delta}V_{ca}$ & ${\Delta}v_{avg}$ & ${\Delta}v_{max}$ &  ${\Delta}e
_{avg}$  &  ${\Delta}e_{max}$    &R$_{eq}$       \\
       & (Hipparcos)& Type &(M$_{\odot}$)&      (pc)            &   (km s$^{-1}$)          & ($10^{-3}$ m s$^{-1}$)& ($10^{-3}$ m s$^{-1}
$) &    ($10^{-3}$)  &   ($10^{-3}$) & (pc) \\\hline
\\
\vspace{0.2cm}
1 & HIP 10798 &G5V& 0.9 & 1.88$^{+0.51}_{-0.40}$ &   42.8$\pm 1.3$   & 33.07$\pm30.61$  &  90.52$\pm69.02$  &   0.54$\pm 0.45$ & 1.47$\pm
0.92$ & 1.09$^{+0.30}_{-0.23}$    \\
\vspace{0.2cm}
2 & HIP 17378 &K0IV& 0.8 & 3.96$^{+0.42}_{-0.41}$&   47.9$\pm 1.5$   &  4.55$\pm 0.96$  &   5.19$\pm 1.17$  &   0.07$\pm 0.02$ & 0.08$\pm
 0.02$ & 2.35$^{+0.25}_{-0.24}$ \\
\vspace{0.2cm}
3 & HIP 19893 &F4III& 1.5 & 4.94$^{+0.19}_{-0.14}$&   13.4$\pm 1.4$   & 18.86$\pm 2.19$  &  20.85$\pm 2.45$  &  0.31$\pm 0.04$ & 0.34$\pm
 0.04$ & 2.55$^{+0.10}_{-0.07}$        \\
\vspace{0.2cm}
4 & HIP 19921 &K2IVa& 0.7& 3.74$^{+0.31}_{-0.33}$&   28.6$\pm 0.6$   &  7.13$\pm 1.15$  &   8.16$\pm 1.41$  &   0.12$\pm 0.02$ & 0.13$\pm
 0.02$ & 2.28$^{+0.19}_{-0.20}$ \\  
\vspace{0.2cm}
5 & HIP 22122 &K1V& 0.8 &       2.76$^{+0.56}_{-0.18}$&   62.7$\pm 1.5$ &  6.08$\pm 1.27$  &   7.26$\pm 1.63$  &   0.10$\pm 0.02$ & 0.12$
\pm 0.03$ &1.64$^{+0.33}_{-0.11}$       \\  
\vspace{0.2cm}
6 & HIP 23693 &F7V& 1.4 &       0.92$^{+0.13}_{-0.12}$&   21.6$\pm 1.5$   & 308.70$\pm97.73$  &  621.14$\pm344.07$  &   5.03$\pm 1.59$ &
10.12$\pm 5.60$ & 0.48$^{+0.07}_{-0.06}$ \\  
\vspace{0.2cm}
7 & HIP 25544 &G6/G8V& 0.9      &       1.49$^{+0.39}_{-0.30}$&   25.5$\pm 0.9$   &56.59$\pm24.86$  &  81.96$\pm42.86$  &   0.92$\pm 0.40
$ & 1.33$\pm 0.70$ & 0.86$^{+0.23}_{-0.17}$\\ 
\vspace{0.2cm}
8 & HIP 27628 &K2III& 0.7       &       0.58$^{+0.51}_{-0.11}$&   83.9$\pm 2.1$   &70.82$\pm47.74$  &  719.08$\pm403.24$  &   1.15$\pm 0.
78$ &11.71$\pm7.04$ & 0.35$^{+0.31}_{-0.07}$\\ 
\vspace{0.2cm}
9 & HIP 29568  &G5V& 0.9        &       2.95$^{+0.90}_{-0.32}$&   11.6$\pm 0.5$   &30.09$\pm 7.77$  &  35.36$\pm 9.88$  &   0.49$\pm 0.13
$ & 0.58$\pm 0.16$ & 1.71$^{+0.52}_{-0.19}$     \\ 
\vspace{0.2cm}
10 & HIP 29958 &M0IV& 0.5       &       1.00$^{+2.36}_{-0.40}$&  102.1$\pm 5.6$   & 7.42$\pm5.27$  &  14.98$\pm11.52$  &   0.12$\pm 0.10$
 & 0.24$\pm 0.17$ & 0.65$^{+1.53}_{-0.26}$      \\ 
\vspace{0.2cm}
11 & HIP 31711 &G1.5V& 1.0      &       3.97$^{+0.37}_{-0.31}$&   14.6$\pm 1.4$   &18.37$\pm 3.26$  &  20.93$\pm 3.90$  &   0.30$\pm 0.05
$ & 0.34$\pm 0.06$ & 2.25$^{+0.21}_{-0.18}$     \\  
\vspace{0.2cm}
12 & HIP 37504 &K0III& 0.8      &       4.59$^{+1.52}_{-0.91}$&   37.6$\pm 1.5$   & 4.39$\pm 2.71$  &   4.99$\pm 3.48$  &   0.07$\pm 0.04
$ & 0.08$\pm 0.06$ & 2.72$^{+0.90}_{-0.54}$     \\ 
\vspace{0.2cm}
13 & HIP 38908 &G0V& 1.1        &       1.97$^{+0.69}_{-0.42}$&   49.5$\pm 0.8$   &22.07$\pm 8.03$  &  29.09$\pm12.35$  &   0.36$\pm 0.13
$ & 0.47$\pm 0.20$ & 1.09$^{+0.38}_{-0.23}$     \\ 
\vspace{0.2cm}
14 & HIP 83990 &K2V& 0.7        &       3.92$^{+1.18}_{-1.16}$&   60.4$\pm 1.5$   & 3.68$\pm 2.49$  &   4.33$\pm 3.72$  &   0.06$\pm 0.04
$ & 0.07$\pm 0.06$ & 2.39$^{+0.72}_{-0.71}$\\
\vspace{0.2cm}
15 & HIP 89042 &G0V& 1.1        &       2.29$^{+1.21}_{-0.21}$&   50.6$\pm 1.6$   &12.14$\pm 4.44$  &  14.84$\pm 5.90$  &   0.20$\pm 0.07
$ & 0.24$\pm 0.10$ & 1.27$^{+0.67}_{-0.12}$     \\
\vspace{0.2cm}
16 & HIP 93506 &A2III& 2.5      &       3.50$^{+2.31}_{-1.21}$&   39.5$\pm 5.7$   & 16.62$\pm12.66$  &  19.37$\pm16.94$  &   0.27$\pm 0.2
1$ & 0.32$\pm 0.28$ &1.58$^{+1.04}_{-0.55}$     \\ 
\vspace{0.2cm}
17 & HIP 114996 &F1III& 1.5     &       3.94$^{+1.35}_{-1.77}$&   29.3$\pm 0.8$   & 25.42$\pm13.03$  &  49.78$\pm28.11$  &   0.41$\pm 0.2
3$ & 0.81$\pm 0.47$ &  2.03$^{+0.70}_{-0.91}$   \\
\vspace{0.2cm}
18 & HIP 116250 &G2V& 1.0       &       2.79$^{+2.39}_{-1.92}$&   22.9$\pm 1.7$   &33.64$\pm23.65$  &  55.80$\pm41.51$  &   0.55$\pm 0.37
$ & 0.91$\pm 0.73$ &  1.58$^{+1.35}_{-1.09}$    \\
\hline
\end{tabular}
\end{table}

\begin{table}
\caption{Maximum number of possible perturber candidates from the $Hipparcos$ Catalog}
\begin{tabular}[h]{ccccccccccc}
\hline\hline
D$_{ca}$ (pc) :        & $<$ 0.2 & $<$ 0.4 & $<$ 0.6 & $<$ 0.8 & $<$ 1 & $<$ 1.2 & $<$ 1.4 & $<$ 1.6 & $<
$ 1.8 & $<$ 2 \\ \hline
t(D$_{ca}$)$>$-100       & 0 & 0 & 0 & 0 & 0 & 0 & 0 & 0 & 0 & 0 \\ 
t(D$_{ca}$)$>$-200       & 0 & 0 & 0 & 0 & 0 & 1 & 1 & 2 & 2 & 2 \\ 
t(D$_{ca}$)$>$-300       & 0 & 0 & 0 & 0 & 2 & 3 & 3 & 4 & 4 & 5 \\ 
t(D$_{ca}$)$>$-400       & 0 & 0 & 1 & 1 & 5 & 7 & 8 & 10 & 10 & 11 \\ 
t(D$_{ca}$)$>$-500       & 1 & 2 & 3 & 6 & 10 & 14 & 20 & 22 & 24 & 25 \\ 
t(D$_{ca}$)$>$-600       & 3 & 5 & 11 & 16 & 24 & 28 & 39 & 41 & 44 & 45 \\ 
t(D$_{ca}$)$>$-700       & 5 & 12 & 21 & 29 & 43 & 49 & 64 & 68 & 81 & 84 \\ 
t(D$_{ca}$)$>$-800       & 12 & 23 & 33 & 45 & 64 & 73 & 91 & 96 & 111 & 119 \\ 
t(D$_{ca}$)$>$-900       & 14 & 27 & 40 & 56 & 77 & 90 & 110 & 117 & 138 & 148 \\ 
t(D$_{ca}$)$>$-1000      & 20 & 36 & 51 & 70 & 93 & 111 & 133 & 141 & 164 & 174 \\ 
\hline 
\end{tabular}
\end{table}

\begin{table}
\caption{Perturber candidates for future radial velocity studies}
\begin{tabular}[h]{lccccc}
\hline\hline
Name     &   D$_{ca}$ & t$_{ca}$&      R$_v$  & [R$_v$](D$_{ca}<$0.5 pc) & $\Delta$V$_{ca}$\\
(Hipparcos)& (pc)     & (kyr)  &      (km s$^{-1}$) & (km s$^{-1}$) & (km s$^{-1}$) \\ \hline
HIP 1837* &      0.025 & -537.0 &         30.6 &  [  29.5,  31.7] &       37.7 \\ 
HIP 6343 &      0.325 & -959.0 &         89.9 &  [  87.8,  92.1] &       78.8 \\ 
HIP 6913 &      0.229 & -560.0 &         88.2 &  [  84.8,  91.9] &      100.9 \\ 
HIP 11324 &     0.127 & -998.0 &         39.7 &  [  38.9,  40.7] &       28.9 \\ 
HIP 11558 &     0.374 & -680.0 &         40.8 &  [  40.2,  41.5] &       41.6 \\ 
HIP 12960* &     0.007 & -732.0 &        103.7 &  [  99.1, 108.7] &       90.6 \\ 
HIP 14007 &     0.327 & -910.0 &         50.3 &  [  49.1,  51.5] &       34.9 \\ 
HIP 14116*&     0.064 & -799.0 &         91.1 &  [  87.1,  95.3] &       74.7 \\ 
HIP 22817* &     0.082 & -725.0 &         98.8 &  [  96.3, 101.2] &       97.5 \\ 
HIP 26067 &     0.490 & -283.0 &        110.1 &  [ 105.0, 116.1] &       90.3 \\ 
HIP 30260 &     0.480 & -908.0 &         58.1 &  [  56.8,  59.4] &       37.6 \\ 
HIP 33973* &     0.077 & -696.0 &         80.2 &  [  77.8,  82.7] &       63.3 \\ 
HIP 42215 &     0.154 & -451.0 &         78.7 &  [  75.9,  81.8] &       65.2 \\ 
HIP 45599 &     0.248 & -897.0 &         71.1 &  [  69.4,  73.0] &       58.0 \\ 
HIP 51578 &     0.431 & -924.0 &         88.8 &  [  87.2,  90.5] &       77.6 \\ 
HIP 57949 &     0.237 & -635.0 &         50.6 &  [  49.2,  52.0] &       61.3 \\ 
HIP 58289 &     0.221 & -403.0 &         91.0 &  [  88.8,  93.4] &      101.6 \\ 
HIP 72455 &     0.273 & -697.0 &        100.6 &  [  98.2, 103.0] &      101.3 \\ 
HIP 78074 &     0.139 & -792.0 &         59.7 &  [  55.1,  65.0] &       79.5 \\ 
HIP 79789* &     0.096 & -871.0 &         80.1 &  [  77.9,  82.3] &       86.9 \\ 
HIP 87435 &     0.110 & -773.0 &         92.2 &  [  88.4,  96.5] &      112.9 \\ 
HIP 104493 &    0.398 & -609.0 &         70.1 &  [  68.8,  71.4] &       84.8 \\ 
\hline
\end{tabular}
\end{table}

\begin{center}
\begin{figure}
	\includegraphics[width=6in]{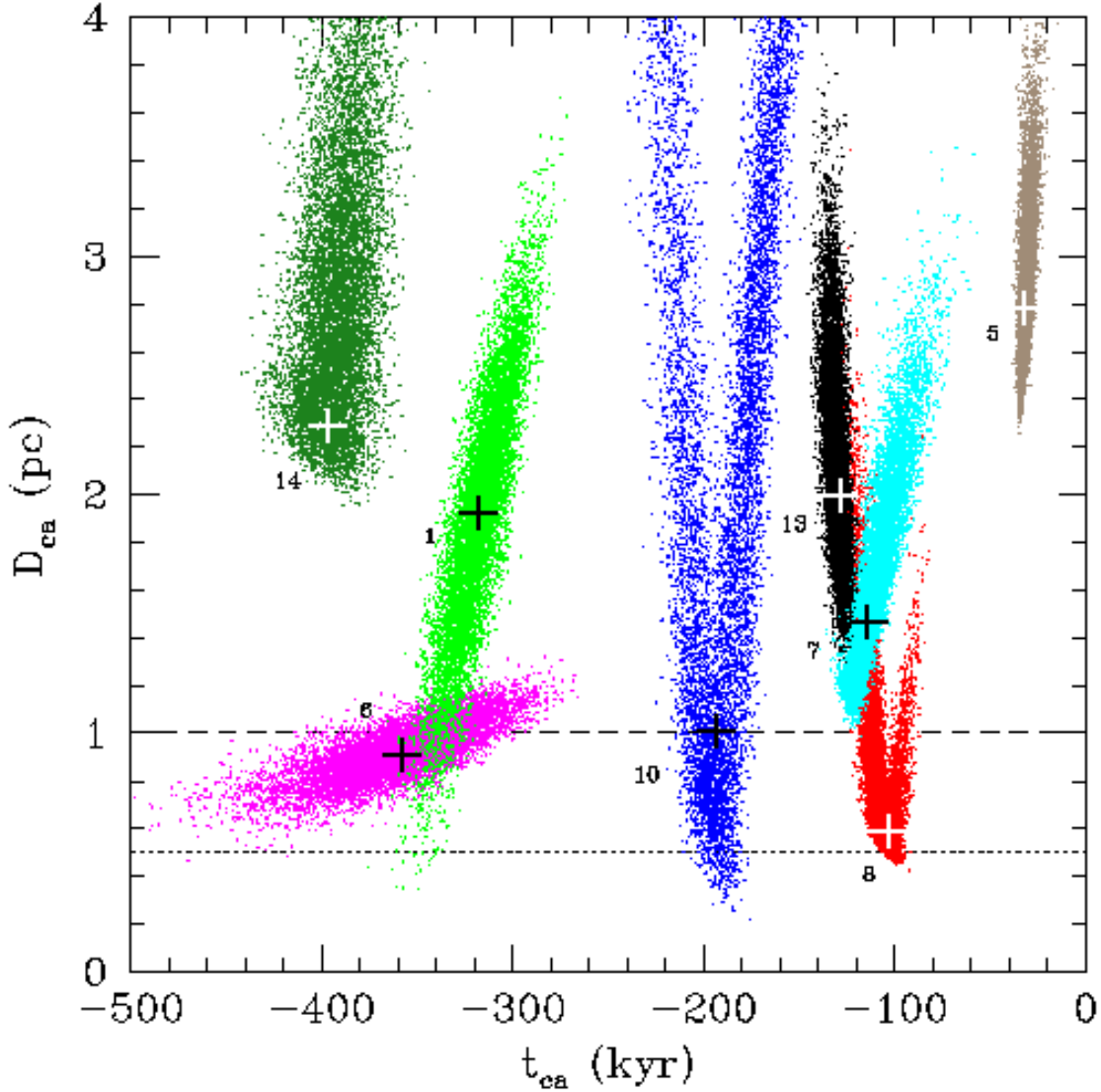}
	\caption{ Positions in the {\it closest approach plane}  ($D_{ca},t_{ca}$) 
from the random Monte-Carlo draw. 
From the 18 candidate perturbers, we show a subsample  
of 8 stars that have ${D}_{ca}<3$ pc. 
Each color corresponds to a different star, which 
have number labels corresponding to column 1 in Table 2 and Table 3.  
Crosses mark the maxima of the probability distributions for (${D}_{ca}$,${t}_{ca}$).
The spread in the 
distribution of points
reflects the initial uncertainties on the stellar proper motions, parallaxes,
and radial velocities.  The horizontal dashed line marks $D$ = 1.0 pc
($\simeq \beta$ Pic's Roche radius)
and the dotted line marks $D$ = 0.5 pc.
}
\end{figure}
\end{center}
 
\begin{figure}
\centering
\subfigure[]{%
	\label{fig:graphics}
	\includegraphics[width=7in]{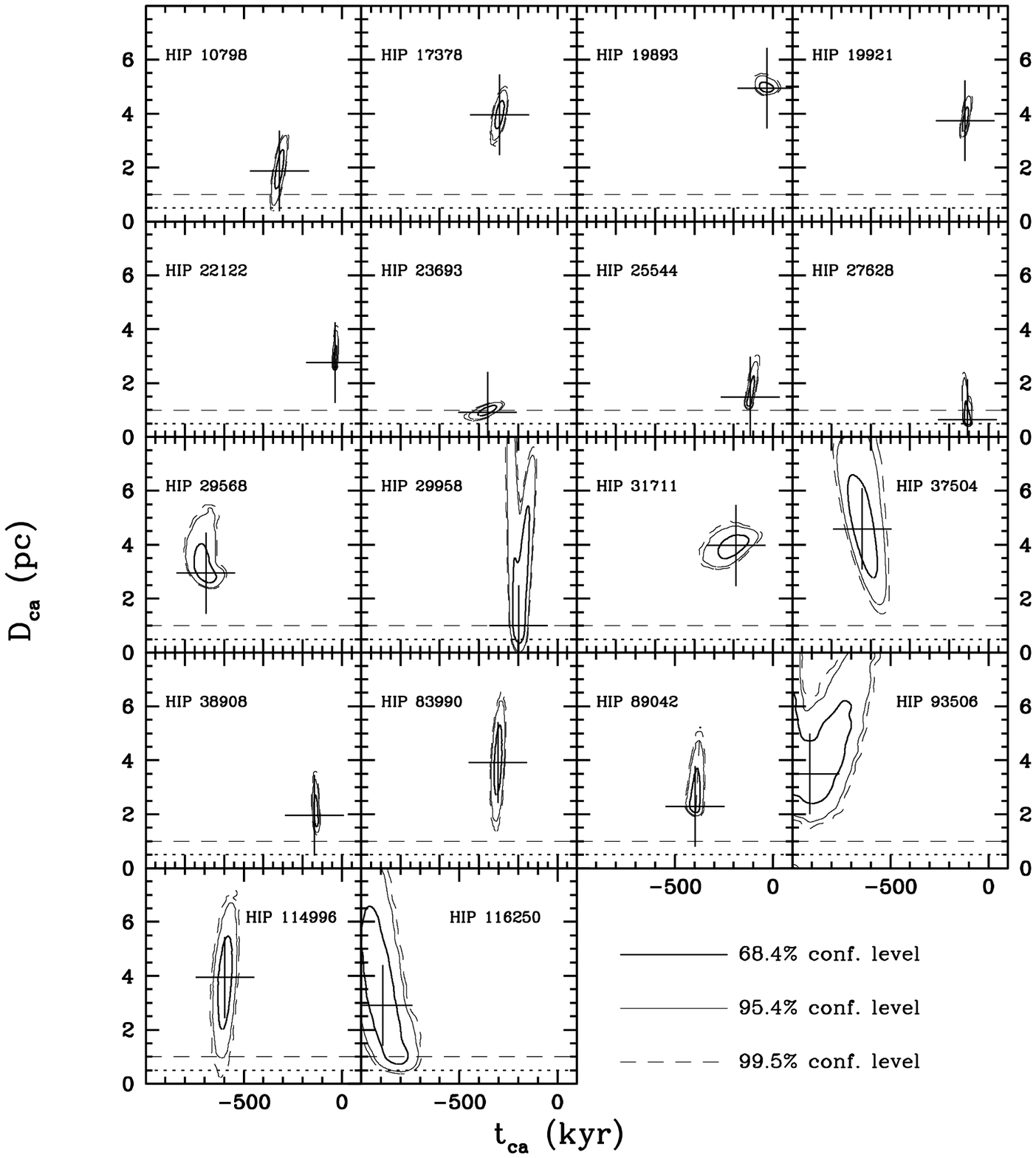}}%
\caption{Figure 2b on next page.}%
\label{fig:graphics}
\end{figure}

\addtocounter{figure}{-1}
\begin{figure}
\addtocounter{subfigure}{1}
\centering
\subfigure[]{%
	\label{fig:graphics}
        \includegraphics[width=6.5in]{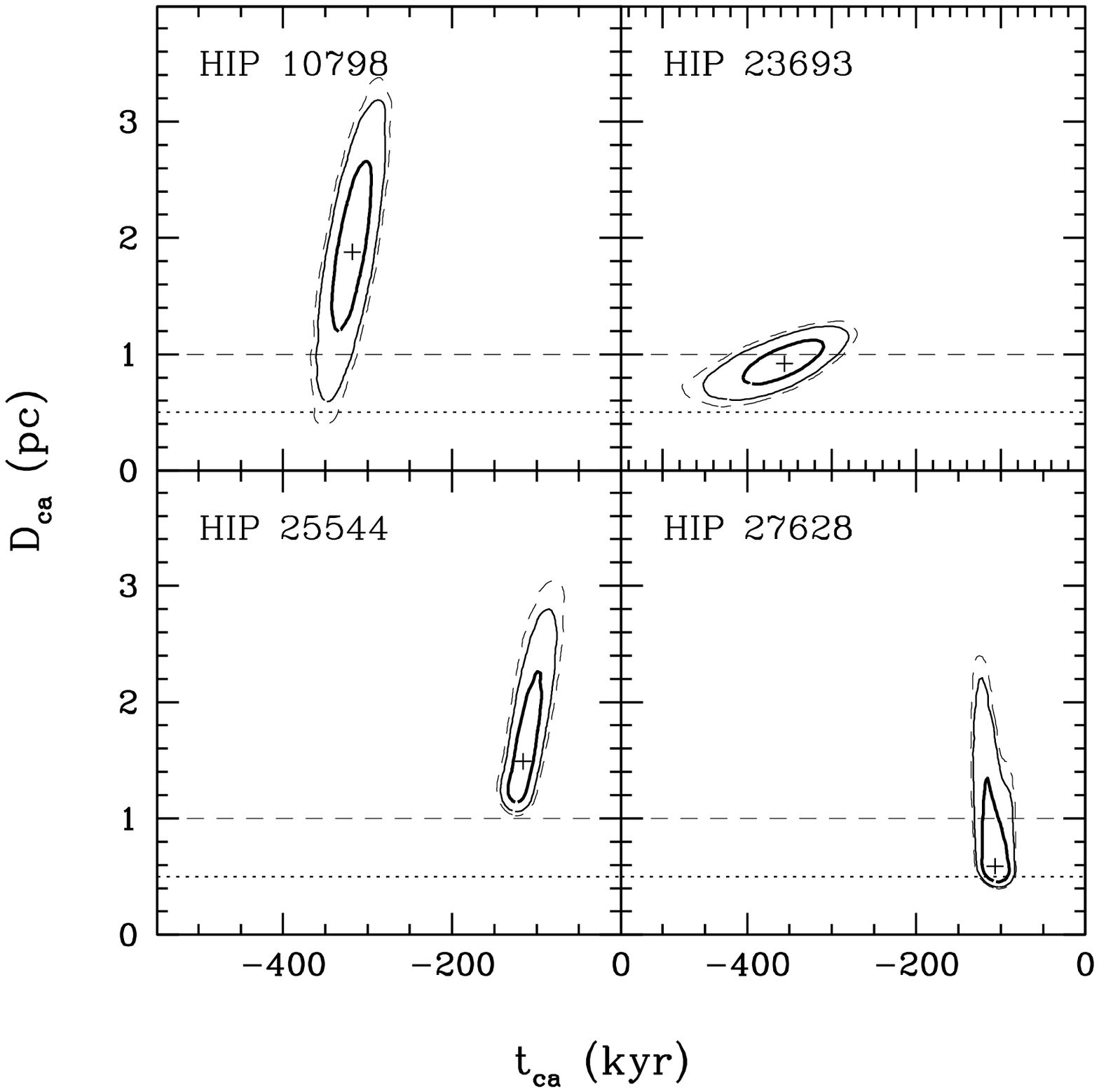}}%
\caption{
Isocontours for the Monte-Carlo distributions 
in the {\it closest approach plane}  (${D}_{ca},t_{ca}$) for the 18 stars with 
${D}_{ca}<5$ pc.  
Bold contour: 68.3$\%$ confidence level, thin contour: 
95.4$\%$, and dashed contour: 99$\%$  confidence level. 
The contours outline the decreasing
probability of finding the star at a more specific location at a given
time due
to uncertainties in proper motion, radial velocity, and parallax.  
Each cross marks the 
position of the maximum of the
probability density distribution in the Monte Carlo simulation (the
size of the crosses has no significance). 
The probabilities that are given in columns 5 and 6 of Table 2
are calculated by dividing the area for each contour below the 1 pc and 0.5 pc
lines by the total area enclosed by the contour.
{\bf Figure 2b} is an enlarged version of {\bf Figure 2a} for the four candidate
perturbers with the smallest ${D}_{ca}$.
}%
\end{figure}

\begin{figure}
\begin{center}
\epsfig{file=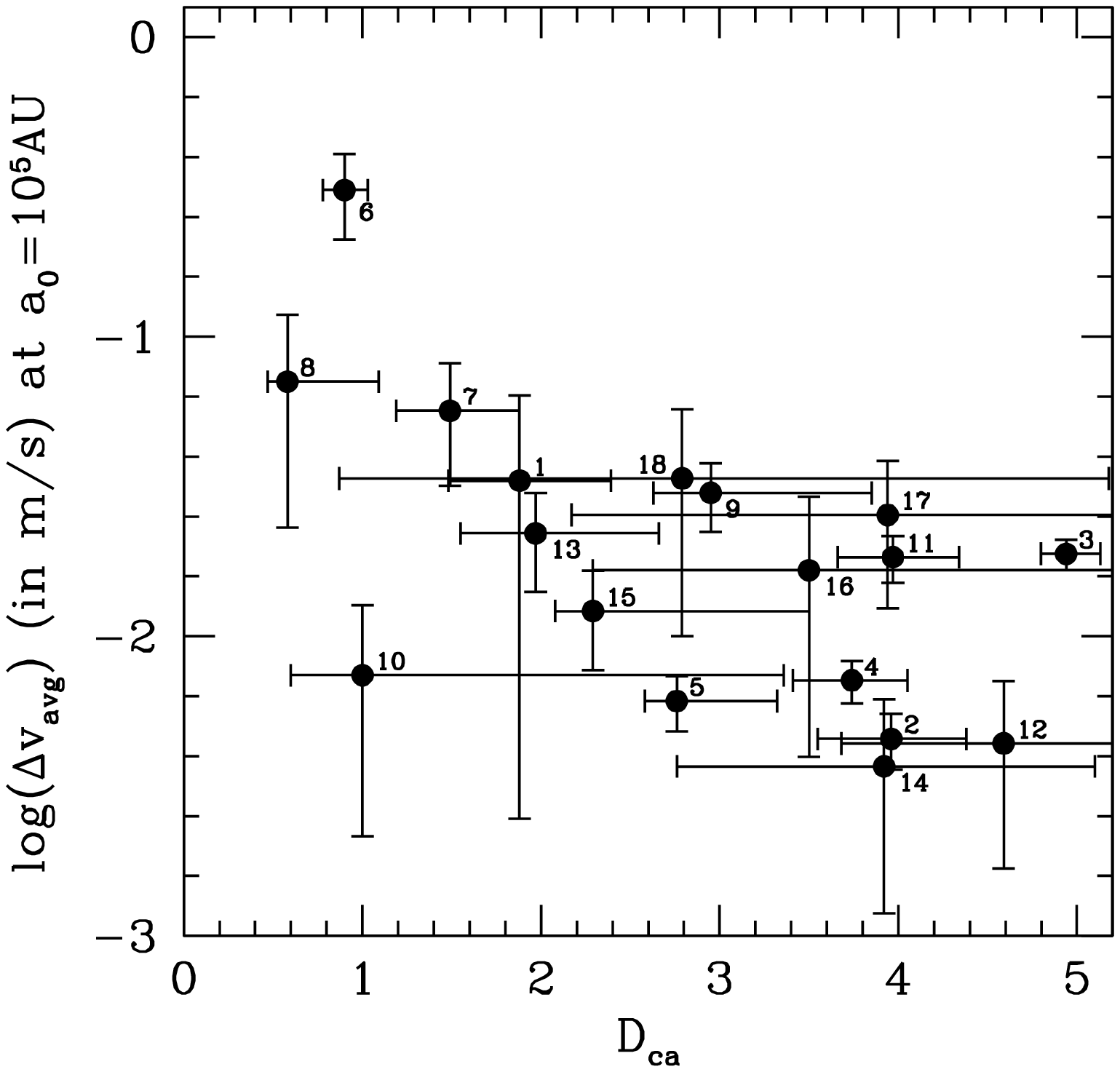,height=9cm,width=9cm}
\epsfig{file=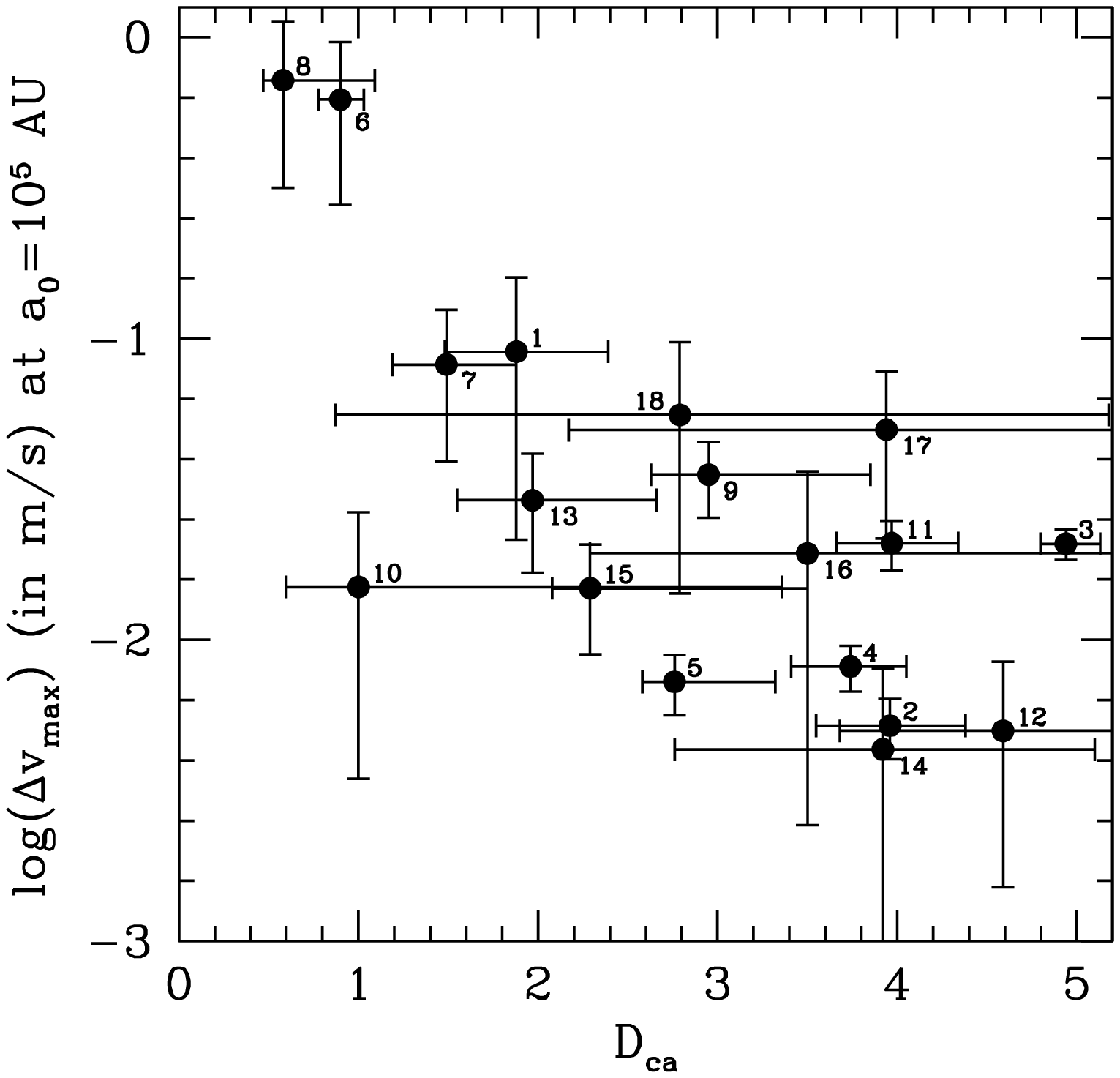,height=9cm,width=9cm}
\caption{
Velocity changes induced by the candidate perturbers on 
hypothetical objects orbiting $\beta$ Pic with semi-major axis $10^5$
AU (equation 1 and data from Table 3).  Fig. 3a (top) shows the mean velocity change
for particles distributed in a shell with radius $10^5$ AU.  Fig. 3b (bottom) plots
the maximum velocity change induced by a given stellar encounter (i.e. for
objects that lie closest to the perturbing star).  Error bars are 1 $\sigma$
(68.4\% confidence level).
Stars are identified with numbering given in 
Tables 2 and 3.
}
\end{center}
\end{figure}

\begin{figure}
\begin{center}
\epsfig{file=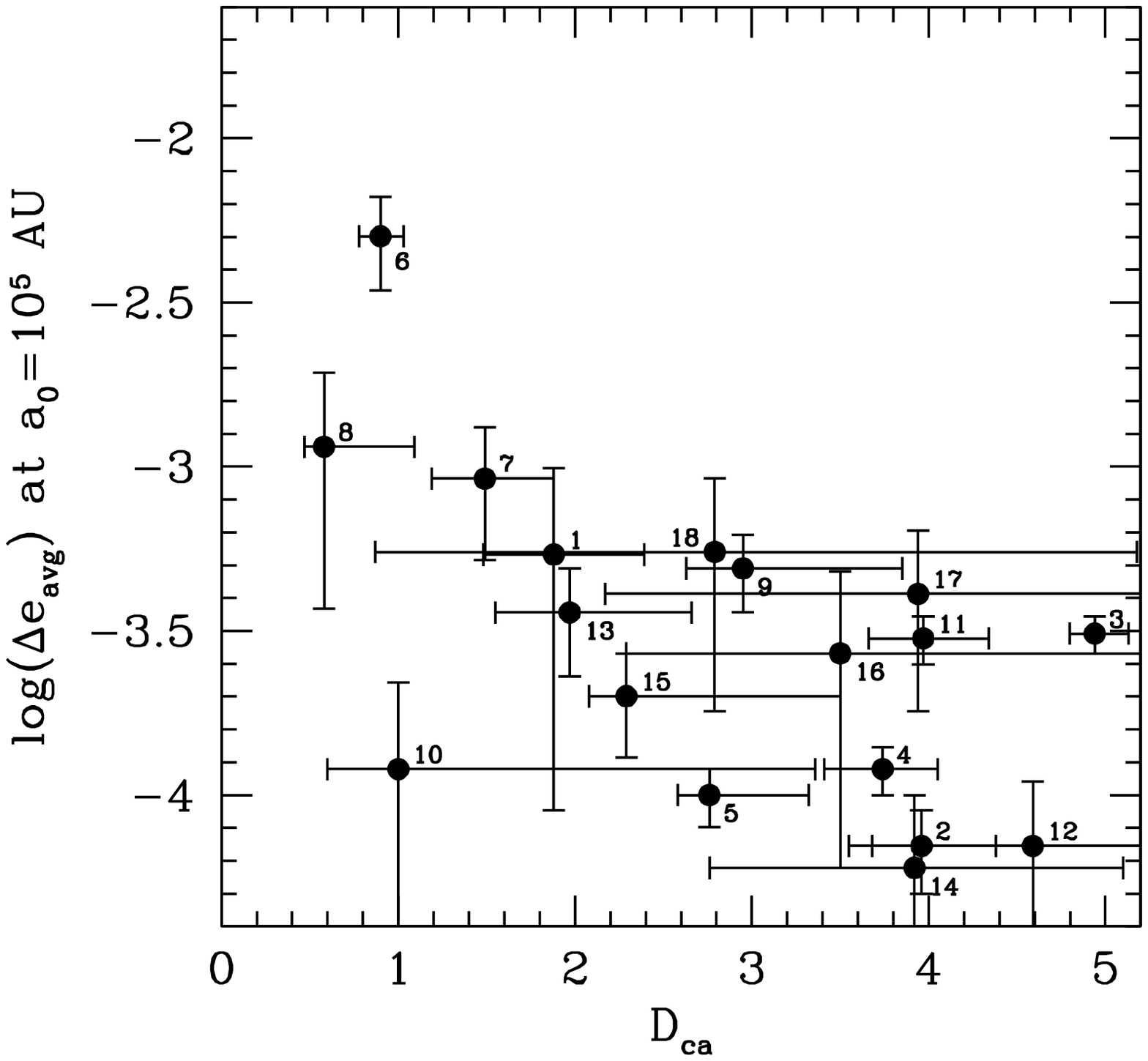,height=9cm,width=9cm}
\epsfig{file=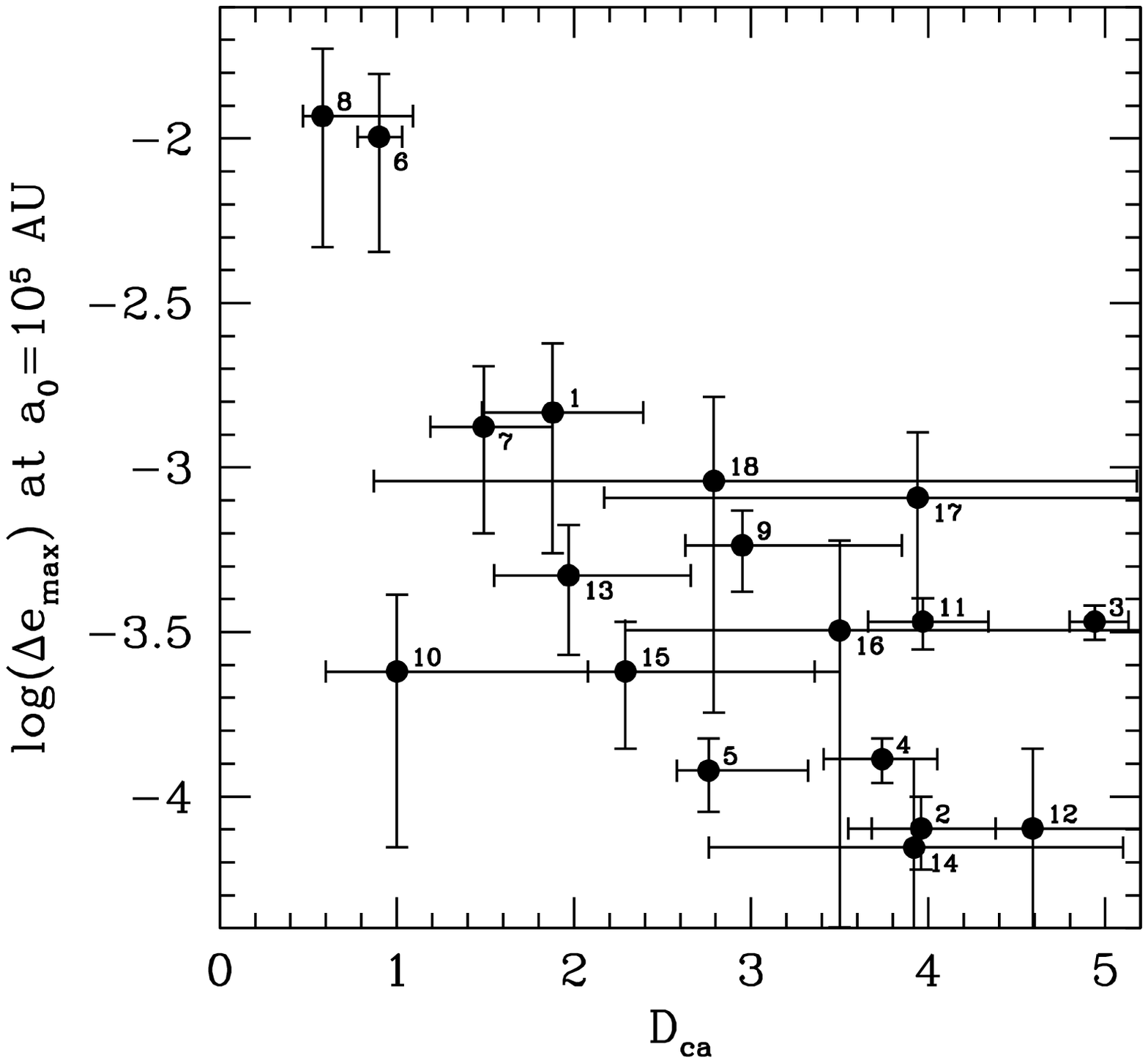,height=9cm,width=9cm}
\caption{
Eccentricity changes induced by the candidate perturbers on a  
hypothetical objects orbiting $\beta$ Pic with semi-major axis $10^5$
AU.  Fig. 4a (top) shows the mean eccentricity change
for particles distributed in a shell with radius $10^5$ AU.  Fig. 4b (bottom) plots 
the maximum eccentricity change induced by a given stellar encounter (i.e. for
objects that lie closest to the perturbing star).  Error bars are 1 $\sigma$
(68.4\% confidence level).  Stars are identified with numbering given in 
Tables 2 and 3.
}
\end{center}
\end{figure}

\begin{center}
\begin{figure}
\includegraphics[width=6.5in]{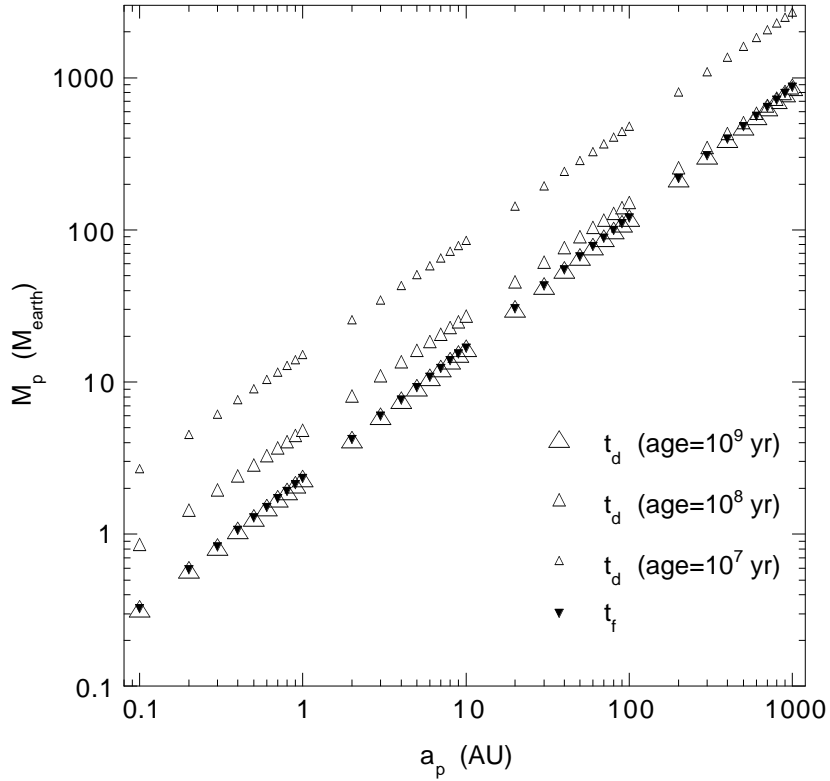}
        \caption{ 
Relationship between planet parameters and Oort cloud formation
timescales.  
Open triangles plot the constraint for the diffusion timescale, $t_d$, assuming 
three different stellar ages (Eqn. 4). The $t_d$ triangles point upwards to indicate the r
egions of
planet parameter space
capable of diffusing comets to the 
$\beta$ Pic Roche radius within the specified age ($t_{d} < t_{\star}$).
Below this boundary, there has been insufficient time to produce a cloud
of comets with large apastra ($\sim 10^5$ AU).
The timescale for freezing the semi-major axis of a comet, $t_f$, is shown by
downward-pointing solid triangles (Eqn. 5).  Below the solid triangles the galactic
tidal field decouples the comets from the planet region before ejection energy
is reached ($t_{f} \leq t_{d}$).  
Comets are ejected if the planet parameters lie above the solid
triangles.  A mature Oort cloud can form around $\beta$ Pic only after 10$^9$ yr
when $t_{f} \leq t_{d} < t_{\beta Pic}$.
}
\end{figure}
\end{center}


\newpage
\begin{thebibliography}{}
\bibitem[Allen \&  Santill\'an(1991)]{als91}
Allen, C. \& Santill\'an, A. 1991, Rev. Mexicana Astron. Astrof., 22, 255
 
\bibitem[Allen \&  Santill\'an(1993)]{als93}
Allen, C. \& Santill\'an, A. 1993, Rev. Mexicana Astron. Astrof., 25, 39

\bibitem[Andersen et al.(1985)]{and85}
Andersen, J., Nordstrom, B., Ardeberg, A., et al. 1985, \aaps, 59, 15

\bibitem[Arymowicz(1994)]{art94}
Artymowicz, P. 1994, in Circumstellar Dust Disks and Planet Formation,
ed. R. Ferlet \& A. Vidal-Madjar (Paris: Editions Frontieres)

%
%

\bibitem[Backman \& Paresce(1993)]{bap93} 
Backman, D. E. \& Paresce, F. 1993,
        in Protostars and Planets III, ed. E. H. Levy \& J. I. Lunine, 
        Tucson: Univ. Arizona Press, 1253


\bibitem[Barrado y Navascues et al.(1999)]{bar99}
Barrado y Navascues, D., Stauffer, J.R., Song, I. and Caillault, J.-P. 1999,
\aap, 520, L123

\bibitem[Barbier-Brossat \& Figon(2000)]{bbf00}
Barbier-Brossat, M. \& Figon, P. 2000, \aaps, 142, 217

\bibitem[Beust \& Morbidelli(2000)]{beu00}
Beust, H. \& Morbidelli, A. 2000, Icarus, 143, 170

\bibitem[Brunini \& Fernandez(1996)]{bru96}
Brunini, A. \& Fernandez, J.A. 1996, \aap, 308, 988

\bibitem[Burrows et al.(1995)]{bur95}
Burrows, C.J., et al. 1995, B.A.A.S., 27, 1329

\bibitem[Cox (1999)]{cox99}
Cox, A. 1999, Allen's Astrophysical Quantities, AIP Press, New York 

\bibitem[Crifo et al.(1997)]{cri97}
Crifo, F., Vidal-Madjar, A., Lallement, R., Ferlet, R. \& Gerbaldi, M.
1997, \aap, 320, L29 

\bibitem[de Boer et al.(1997)]{deb97}
de Boer et al. 1997, \aap, 327, 577

\bibitem[De Medeiros \& Mayor(1999)]{dem99}
De Medeiros, J.R. \& Mayor, M. 1999, \aaps, 139, 433
 
\bibitem[Duflot et al.(1995)]{duf95}
Duflot, M., Fignon, P., Meyssonnier, N. 1995, \aaps, 114, 269

\bibitem[ESA(1997)]{esa97}
ESA, 1997, The Hipparcos and Tycho Catalogues, ESA SP-1200,
Noordwijk, ESA

\bibitem[Fernandez(1997)]{fer97}
Fern\'andez, J.A. 1997, Icarus, 129, 106

\bibitem[Fernandez \& Ip(1991)]{fip91}
Fern\'andez, J.A. \& Ip, W.-H. 1991, in Comets in the Post-Halley Era, 
eds R.L. Newburn et al., Kluwer, Dordrecht, p. 487


\bibitem[Garc\'{\i}a-S\'anchez et al.(1999)]{gar99}
Garc\'{\i}a-S\'anchez, J., Preston, R. A., Jones, D. A., et al.  1999, \aj, 117, 1042

\bibitem[Goldreich \& Tremaine(1980)]{gol80}
Goldreich, P. \& Tremaine, S. 1980, \apj, 241, 425

\bibitem[Greaves et al.(1998)]{gre98}
Greaves, J.S., Holland, W.S., Moriarty-Schieven, G. et al. 1998, \apjl, 506, L133

\bibitem[Grenier et al.(1999)]{gre99}
Grenier, S., Burnage, R., Faraggiana, R., et al. 1999, \aaps, 135, 503


\bibitem[Heap et al.(2000)]{hea00} 
Heap, S.R., Lindler, D.J., Lanz, T.M.,  et al.
2000, \aj, 539, 435

\bibitem[Henney \& O'Dell(1999)]{hen99}
Henney, W.J. \& O'Dell, C.R. 1999, \aj, 118, 2350

\bibitem[Holmberg \& Flynn(2000)]{hol00}
Holmberg, J. \& Flynn, C. 2000, \mnras, 313, 209

\bibitem[Johnson \& Soderblom(1987)]{joh87}
Johnson, D.R.H. \& Soderblom, D.R. 1987, \aj, 98, 864

\bibitem[Kalas et al.(2000)]{kal00} 
Kalas, P., Larwood, J.D., Smith, B.A., \& Shultz, A.
2000, \apjl, 530, 133

\bibitem[Kalas \& Jewitt(1995)]{kal95} Kalas, P. \& Jewitt, D.  1995, \aj,
    110, 794
 
\bibitem[Lagage \& Pantin(1994)]{lag94}
Lagage, P.O. \& Pantin, E. 1994, Nature, 369, 628

\bibitem[Lagrange et al.(1995)]{lag95}
Lagrange, A.-M., Vidal-Madjar, A., Deleuil, M., Emerich, C., Beust, H.,
\& Ferlet, R. 1995, \aap, 296, 499

\bibitem[Larwood(1997)]{lar97}
Larwood, J.D. 1997, \mnras, 290, 490

\bibitem[Larwood \& Kalas(2001)]{lar00}
Larwood, J.D. \& Kalas, P. 2001, \mnras, in press

\bibitem[Laughlin \& Adams(1998)]{lau98}
Laughlin, G. \& Adams, F.C. 1998, \apjl, 508, L171

\bibitem[Lecavelier des Etangs et al.(1997)]{lec97}
Lecavelier des Etangs, A., Vidal-Madjar, A., Burki, G.  et al.
1997, \aap, 328, 311

\bibitem[Liou \& Zook(1999)]{liu99}
Liou, J.-C. \& Zook, H.A. 1999, \aj, 118, 580

\bibitem[Malhotra et al.(2000)]{mal00}
Malhotra, R., Duncan, M.J. \& Levison, H.F.
2000, in Protostars and Planets IV, ed. V. Mannings, A. P. Boss \& S. S.
Russell (Tucson: University of Arizona Press), in press


\bibitem[Marcy, Butler \& Vogt(2000)]{mbv00}
Marcy, G.W., Butler, R.P. \& Vogt S.S. 2000, \apjl, 536, L43
 
\bibitem[Marcy et al.(2000)]{mar00}
Marcy, G.W., Cochran, W.D., \& Mayor, M. 
2000, in Protostars and Planets IV, ed. V. Mannings, A. P. Boss \& S. S.
Russell (Tucson: University of Arizona Press), in press

\bibitem[Matthews(1994)]{mat94}
Matthews, R.A.J  1994, Q. J. R. astr. Soc., 35, 1

\bibitem[Miyamoto \& Nagai(1975)]{miy75}
Miyamoto, M. \& Nagai, R. 1975, PASJ, 27, 533

\bibitem[Mouillet et al.(1997)]{mou97} 
Mouillet, D., Larwood, J. D., Papaloizou, J. C. B., and 
Lagrange, A-.M.  1997, \mnras, 292, 896  

\bibitem[M\"ullari \& Orlov(1996)]{mul96}
M\"ullari, A.A. \& Orlov V.V. 1996, Earth, Moon, and Planets, 72, 19

\bibitem[Nordstrom \& Andersen(1985)]{nor85}
Nordstrom, B. \& Andersen, J. 1985, \aaps, 61, 53

\bibitem[Oort(1950)]{oor50}
Oort, J.H. 1950, B.A.N., Vol. XI, pp. 91. 

\bibitem[Pantin et al.(1997)]{pan97}
Pantin, E., Lagage, P.O. \& Artymowicz, P. 1997, \aap, 327, 1123

\bibitem[Roques et al.(1994)]{roq94}
Roques, F., Scholl, H., Sicardy, B. \& Smith, B. 1994, Icarus, 108, 37

\bibitem[Schuster \& Allen(1997)]{sch97}
Schuster, W.J. \& Allen, C. 1997, \aap, 319, 796

\bibitem[Smith \& Terrile(1984)]{smi84}
Smith, B. \& Terrile, R. 1984, Science, 226, 1421

\bibitem[Tremaine(1993)]{tre93}
Tremaine, S.  1993, in {\it Planets Around Pulsars}, ASP Conf. Series, Vol. 35, p. 335


\bibitem[Weissman(1996)]{wei96}
Weissman, P.R. 1996, Earth, Moon, and Planets, 72, 25




\bibitem[Wyatt et al.(2000)]{wya00}
Wyatt, M.C. Dermott, S.F. Telesco, C.M., Fisher, R.S., Grogan, K., Holmes, E.K. \& Pina, K.
2000, \apj, 527, 918

\noindent
\end{thebibliography}
\end{document}